\documentclass[preprintnumbers,amssymb, nofootinbib, nobibnotes,aps,prd]{revtex4}
\usepackage{soul, color}
\usepackage{amsmath}

\usepackage{graphicx}
\usepackage{dcolumn}
\usepackage{bm}

\newcommand{\beq}{\begin{equation}}
\newcommand{\eeq}{\end{equation}}

\newcommand{\p}{\phi}

\newcommand{\del}{\partial}

\newcommand{\ft}{\tilde{f}}

\newcommand{\J}{\mathcal{J}}

\begin{document}

\title{Classical gluon and graviton radiation from the bi-adjoint scalar double copy}%

\author{Walter D. Goldberger}%
\author{Siddharth G. Prabhu}
\author{Jedidiah O. Thompson}
\affiliation{Physics Department, Yale University, New Haven, CT 06520, USA}
\date{\today}%

\begin{abstract}

We find double copy relations between classical radiating solutions in Yang-Mills theory coupled to dynamical color charges and their counterparts in a cubic bi-adjoint scalar field theory which interacts linearly with particles carrying bi-adjoint charge.   The particular color-to-kinematics replacements we employ are motivated by the BCJ double copy correspondence for on-shell amplitudes in gauge and gravity theories.  They are identical to those recently used to establish relations between classical radiating solutions in gauge theory and in dilaton gravity.   Our explicit bi-adjoint solutions are constructed to second order in a perturbative expansion, and map under the double copy onto gauge theory solutions which involve at most cubic gluon self-interactions.   If the correspondence is found to persist to higher orders in perturbation theory, our results suggest the possibility of calculating gravitational radiation from colliding compact objects, directly from a scalar field with vastly simpler (purely cubic) Feynman vertices.

\end{abstract}

\maketitle

\section{Introduction}
\label{sec:intro}

The close connection between ordinary quantum field theories and gravitational phenomena has been one of the major themes of theoretical physics in the last two decades. While gauge/gravity dualities are usually manifest when the field theory side is strongly coupled and the gravitational dual is weakly interacting (as in the AdS/CFT correspondence) it has more recently become clear due to the work of Bern, Carrasco, and Johansson (BCJ)~\cite{Bern:2008qj} that there exist non-trivial relations between observables even at the level of perturbation theory.  The relations uncovered in~\cite{Bern:2008qj} yield a definite prescription for obtaining scattering amplitudes in perturbative gravity directly from those of a gauge theory, simply by replacing the color structures in the Feynman diagrams by suitably defined kinematic structures.   This double copy mapping~\cite{Bern:2010ue} between gauge and gravity theories contains as a special case the earlier KLT identities~\cite{Kawai:1985xq} for string amplitudes, and has been  established~\cite{Bern:2010yg} in the case of tree-level field theory using modern on-shell amplitude techniques.  Although at the loop level the BCJ relations remain conjectural~\cite{Bern:2010ue}, many explicit non-trivial computational checks exist; see~\cite{Carrasco:2015iwa} for a review of the literature.

A natural question is whether the BCJ double copy structure between gauge and gravity theories relates observables besides the perturbative $S$-matrix.    This was first raised in~\cite{Monteiro:2014cda} which proposed a double copy mapping between non-perturbative classical solutions in pure Yang-Mills theory and Einstein gravity.   See~\cite{luna,related} for related work.    In~\cite{Goldberger:2016iau}, it was found that there is a double copy of perturbative classical radiating solutions in any spacetime dimension $d$.  On the Yang-Mills side, the setup of~\cite{Goldberger:2016iau} consists of a set of radiating color charges, interacting self-consistently by gluon exchange.   Applying color-kinematics substitutions to the long distance radiation field ${\cal A}^{\mu a}$ of this system then yields a well-defined double-copy field ${\hat{\cal A}}^{\mu\nu}$ that precisely matches all radiation channels $(\phi,h_{\mu\nu},B_{\mu\nu})$ in a theory of gravitating point sources.    As in the case of the BCJ relations for scattering amplitudes, the double copy gravitational theory is not pure gravity but rather contains additional fields (as expected by counting on-shell degrees of freedom $\epsilon^a_\mu\rightarrow \epsilon_\mu {\tilde \epsilon}_\nu$) that must couple to the point particle sources.    

Part of the motivation for studying the perturbative double copy of classical solutions is the hope of translating the calculation of precision gravitational wave observables (e.g black hole mergers seen at LIGO and other detectors) to the analogous Yang-Mills problem, whose Feynman diagram expansion is considerably simpler.     The primary obstacle to this program is the need to efficiently remove the unwanted states $\phi,B_{\mu\nu}$ from the double copy in order to reproduce pure Einstein gravity.    In the context of BCJ duality for scattering amplitudes, various procedures have been introduced~\cite{Johansson:2014zca,Bern:2015ooa} for projecting out the contributions of the additional modes in Feynman diagrams.   However, these techniques are operative at the loop level, and it is not yet clear if analogous methods can be applied to the types of off-shell, tree-level Feynman diagrams that arise in the calculations of ref.~\cite{Goldberger:2016iau} (these sort of diagrams were first employed by Duff~\cite{Duff} to reproduce the Schwarzschild solution in perturbation theory, and extended to the radiating (time dependent) two-body problem in general relativity in~\cite{Goldberger:2004jt}).   We also note that recently ref.~\cite{Luna:2016hge} has proposed an approach to the double copy of perturbative classical solutions in which there is more freedom in choosing the field content of the gravitational theory.   The results of ref.~\cite{Luna:2016hge} were carried out up to third perturbative order for static spherically symmetric objects, and it would be interesting to see if this approach can be extended to dynamical sources as well.

In this note, we set aside the issue of canceling the unwanted fields and instead focus on the possibility of further simplification of the Feynman rules needed to construct perturbative classical solutions in gravity.    We follow an observation made in ref.~\cite{Luna:2015paa}, that linearized classical Yang-Mills solutions can be interpreted as double copies of field configurations $\phi^{a{\tilde a}}(x)$ in a massless scalar field theory with cubic interaction, 
\beq
\label{eq:phi3}
f^{abc} {\tilde f}^{{\tilde a}{\tilde b}{\tilde c}}  \phi^{a{\tilde a}} \phi^{b{\tilde b}} \phi^{c{\tilde c}}
\eeq
invariant under a global $G\times {\tilde G}$ symmetry with $\phi^{a{\tilde a}}$ in the bi-adjoint representation.   Ref.~\cite{Cachazo:2013iea} showed that, in $d$ spacetime dimensions, tree-level scattering amplitudes in this particular bi-adjoint scalar field theory map under the double copy (acting on ${\tilde G}$) to pure gluon amplitudes in Yang-Mills.   Similar cubic bi-adjoint structures have been introduced in the context of color-kinematics duality~\cite{otherphi31,otherphi32,otherphi33,otherphi34}.    Bi-adjoint scalars have also been shown to play a role in the double copy between supersymmetric Yang-Mills theories and supergravities in diverse dimensions~\cite{Anastasiou:2014qba}, in color-kinematic relations for scalar effective field theories~\cite{Cheung:2016prv}, and in the context of soft theorems and asymptotic symmetries~\cite{Campiglia:2017dpg}. Non-perturbative static solutions of the (source free) bi-adjoint field equations have been recently constructed in ref.~\cite{White:2016jzc}.

In this paper, we obtain perturbative classical solutions of the bi-adjoint scalar with the interaction Eq.~(\ref{eq:phi3}) coupled to point-like bi-adjoint charges $c^{a}(s)$, ${\tilde c}^{{\tilde a}}(s)$ transforming in the adjoint representations of $G$ and ${\tilde G}$. These sources are not treated as fixed but instead evolve self-consistently in the classical field they collectively generate.  Our focus is on the long distance scalar radiation field generated by a set of interacting bi-adjoint color charges coming in from infinity.    We construct the classical field ${\cal A}^{a{\tilde a}}$ formally, as a momentum space integral involving the initial momenta $p^\mu$ and initial charges $c^{a}, {\tilde c}^{\tilde a}$, as well as the momentum of the outgoing radiation.   By applying color-kinematic substitutions similar to those of BCJ, 
we obtain the double copy field ${\cal A}^a_\mu$ that precisely matches the long distance radiation gluon field in the corresponding system of scattering point like color charges $c^a.$  (This solution has been constructed in ref.~\cite{Kovchegov:1997ke,Gyulassy:1997vt} in the four-dimensional case, and generalized to any spacetime dimension $d$ by~\cite{Goldberger:2016iau}).   

The color-kinematic substitutions are precisely of the same form as those used by~\cite{Goldberger:2016iau} to generate gravitational solutions from Yang-Mills theory.   The results presented here together with those of~\cite{Goldberger:2016iau}  then imply a two-fold double copy of classical solutions,
\beq
{\cal A}^{a{\tilde a}}\mapsto {\cal A}^{a\mu}\mapsto {\cal A}^{\mu\nu}
\eeq
that produces the gravitational radiation field ${\cal A}^{\mu\nu}$ due to a collection of dynamical point-like sources from the simpler bi-adjoint radiation field ${\cal A}^{a{\tilde a}}$.    If this pattern is also found to hold at higher orders in perturbation theory, it would allow the calculation of gravitational radiation observables directly from a theory with cubic vertices, sidestepping the vastly more complex tower of interaction vertices in gravity and streamlining computations.

\section{Gluon radiation and its gravitational double copy}

\label{sec:glue}

We begin by reviewing the $d$-dimensional classical gluon radiation solutions found in~\cite{Goldberger:2016iau}.   We couple the Yang-Mills equations\footnote{The conventions are $D_\mu = \partial_\mu + i g A^a_\mu T^a$,  $[T^a,T^b]=if^{abc} T^c$, $(T_{\mbox{\tiny{adj}}}^a)^b_c=-if_{abc}$.} to point particles $x^\mu(s)$ carrying color charge degrees of freedom~\cite{sikivie} $c^a(s)$ which transform in the adjoint representation:
\begin{equation}
\label{eq:YM}
D_\nu F^{\nu\mu}_a(x) = g J_a^\mu(x),
\end{equation}
where the color current sourced by the moving particles is
\beq
\label{eq:ccurrent}
J^\mu_a(x)=\sum_\alpha \int ds c_{\alpha}^a(s) p^\mu_\alpha(s) \delta^d(x-x_\alpha(s)),
\eeq
with a label $\alpha$ that distinguishes the different point charges carrying momentum $p_\alpha^\mu=dx_\alpha^\mu/ds$. The equations of motion follow as a consequence of conservation laws.    Covariant conservation of  $J_a^\mu(x)$ implies that the charges are parallel transported in color space along the particle worldline, $p\cdot D c^a =0$.   In turn, conservation of energy-momentum yields the non-Abelian Lorentz force law for each particle,
\beq
\label{eq:lorentz}
{d p^\mu\over ds}  = g  c^a F_a^{\mu}{}_{\nu}  p^\nu.
\eeq

Ref.~\cite{Goldberger:2016iau} constructed solutions to these equations corresponding to a set of particles coming in from spatial infinity, with initial conditions $c_\alpha^a(s\rightarrow-\infty) = c_\alpha^a$ and 
\beq
x^\mu_\alpha(s\rightarrow-\infty) = b^\mu_\alpha + p^\mu_{\alpha} s.
\eeq
The main object of interest is the classical radiation field measured at future null infinity ($r=|{\vec x}|\rightarrow\infty$ and fixed retarded time $t$).    It has a simple relation to the conserved (but gauge dependent) current 
\beq
\label{eq:gcurr}
{\tilde J}^\mu_a(x)=J^\mu_a +  f^{abc} A^b_\nu(\partial^\nu A_c^\mu - F_c^{\mu\nu}).
\eeq    
For example, in four spacetime dimensions, the long distance radiation field is related to the on-shell current ${\tilde J}^\mu_a(k) =\int d^d x e^{ik\cdot x} {\tilde J}^\mu_a(x)$, $k^2=0$, by 
\beq
\label{eq:lda}
\lim_{r\rightarrow\infty}\langle  A^a_\mu\rangle(x)  = {g\over 4\pi r}\int {d\omega\over 2\pi} e^{-i\omega t} {\tilde J}^\mu_a(k)
\eeq
with $k^\mu = (\omega,{\vec k})=\omega(1,{\vec x}/r)$ .   Similar expressions hold in general spacetime dimension $d$.

\begin{figure}
\centering
\includegraphics[scale=0.3]{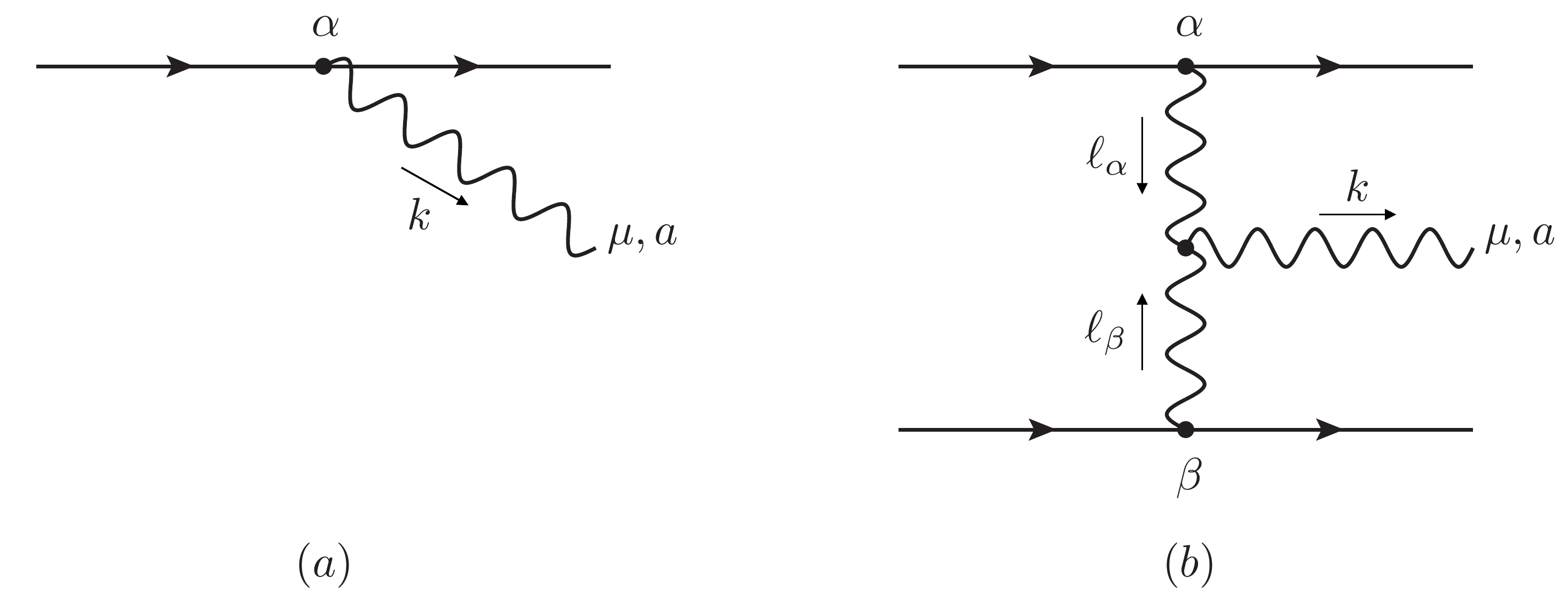}
\caption{Leading order Feynman diagrams for the perturbative expansion of ${\tilde J}^\mu_a(k)$.\label{fig:gluon1pt}}    
\end{figure} 

As long as these particles remain well separated, with sufficiently large impact parameters $b_{\alpha\beta}^\mu=b^\mu_\alpha-b^\mu_\beta$, the on-shell current ${\tilde J}^\mu_a(k)$ can be calculated in perturbation theory.   Up to second order in the gauge coupling, it is given by the Feynman diagrams shown in Fig.~\ref{fig:gluon1pt}.   These diagrams are computed using standard Yang-Mills Feynman rules, with insertions of the classical particle current Eq.~(\ref{eq:ccurrent}).   The leading order result is from Fig.~\ref{fig:gluon1pt}(a) evaluated using static particle trajectories with constant color charge $c^a$ and momentum $p^\mu_\alpha$
\beq
{\tilde J}^\mu_a(k) = \sum_\alpha e^{ik\cdot b_\alpha} c^a_\alpha p^\mu_\alpha (2\pi)\delta(k\cdot p_\alpha)+{\cal O}(g^2).
\eeq
For on-shell $k^2=0$, this is only non-vanishing if $k^\mu$ is along one of the particle momenta $p^\mu_\alpha$, and consequently there is no gluon radiation at this order in perturbation theory, $g\epsilon_\mu^a(k) {\tilde J}^\mu_a(k^2=0)=0$.     At order $g^2$, the deflection of the orbits and color charges due to the Coulomb potential generated by each particle must be taken into account in Fig.~\ref{fig:gluon1pt}(a), which yields a contribution to ${\tilde J}^\mu_a(k)$ of the form
\begin{eqnarray}
\label{eq:ym1}	
\nonumber
\left. {\tilde J}^\mu_a(k)\right|_{\mbox{\footnotesize{Fig.}}~\ref{fig:gluon1pt}(a);{\cal O}(g^2)} &=&  g^2\sum_{\alpha,\beta\atop \alpha\neq\beta} \int_{\ell_\alpha,\ell_\beta} \mu_{\alpha,\beta}(k) {\ell_\alpha^2\over k\cdot p_\alpha} \Bigg[((c_\alpha\cdot c_\beta) c_\alpha)^a \left\{-p_\alpha\cdot p_\beta \left(\ell^\mu_\beta -{k\cdot\ell_\beta\over k\cdot p_\alpha} p^\mu_\alpha\right) +k\cdot p_\alpha p^\mu_\beta - k\cdot p_\beta p^\mu_\alpha\right\}\\
& & \hspace{1cm} {}+ [c_\alpha,c_\beta]^a (p_\alpha\cdot p_\beta) p_\alpha^\mu \Bigg].
\end{eqnarray}
where $[c_\alpha,c_\beta]^a \equiv  i f^{abc} c_\alpha^b c_\beta^c$.  In addition, there is a contribution from the three-gluon vertex, which can be calculated using static paths,
\beq
\label{eq:ym2}
\left. {\tilde J}^\mu_a(k)\right|_{\mbox{\footnotesize{Fig.}}~\ref{fig:gluon1pt}(b);{\cal O}(g^2)}
=  g^2\sum_{\alpha,\beta\atop\alpha\neq\beta} \int_{\ell_\alpha,\ell_\beta}\mu_{\alpha,\beta}(k) [c_\alpha ,c_\beta]^a \left[2(k\cdot p_\beta) p^\mu_\alpha -  (p_\alpha\cdot p_\beta)\ell^\mu_\alpha\right].
\eeq
The total current ${\tilde J}^\mu_a(k)$ is the sum of Eqs.~(\ref{eq:ym1}),~(\ref{eq:ym2}). In these equations, we have defined
\beq
\label{eq:mudef}
\mu_{\alpha,\beta}(k) = 
\left[(2\pi)\delta(p_\alpha\cdot\ell_\alpha)  {e^{i\ell_\alpha\cdot b_\alpha}\over\ell^2_\alpha}\right] \left[(2\pi)\delta(p_\beta\cdot\ell_\beta)  {e^{i\ell_\beta\cdot b_\beta}\over\ell^2_\beta}\right] (2\pi)^d \delta^d(k-\ell_\alpha-\ell_\beta)
\eeq
and momentum integrals\footnote{Here and in what follows, it is implicit that we use retarded boundary conditions $1/k^2 = 1/((k^0+i\epsilon)^2 - {\vec k}^2)$ and $1/k\cdot p=1/(k\cdot p +i\epsilon)$ as is appropriate for classical solutions.} are denoted by $\int_\ell = \int {d^d\ell/(2\pi)^d}$.    This result is given only at the level of the integrand and holds in any dimension $d$.    We note in particular that the term proportional to $f^{abc}$ receives contributions both from the three-gluon interaction and from the time evolution of the color charges, ${\dot c}^a =  g f^{abc} p^\mu A^b_\mu(x(s)) c^c(s)$.

\subsection{Double copy}

Given the result in Eqs.~(\ref{eq:ym1}),~(\ref{eq:ym2}), we may define its gravitational double copy by making color-to-kinematics substitutions.   We replace initial color charges with the respective momenta,
\beq
\label{eq:sub1}
c^a \mapsto p^\mu,
\eeq
the adjoint generators with the three-gluon vertex kinematic structure
\beq
\label{eq:sub2}
i f^{a_1 a_2 a_3}\mapsto \Gamma^{\nu_1\nu_2\nu_3}(q_1,q_2,q_3) =-{1\over 2} \left[\eta^{\nu_1\nu_3}(q_1-q_3)^{\nu_2} +\eta^{\nu_1\nu_2} (q_2-q_1)^{\nu_3} + \eta^{\nu_2\nu_3}(q_3-q_2)^{\nu_1}\right],
\eeq
where $q_1+q_2+q_3=0$, and gluon polarizations by a product of independent polarizations $\epsilon^a_\mu(k)\mapsto \epsilon_\mu(k) {\tilde\epsilon}_\nu(k)$.  This defines a double copy radiation field ${\hat{\cal A}}^{\mu\nu}(k)$, with $k^2=0$, 
\beq
\epsilon^a_\mu(k) {\cal A}_\mu^a(k) \mapsto \epsilon_\mu(k) {\tilde\epsilon}_\nu(k) {\hat{\cal A}}^{\mu\nu}(k),
\eeq
where ${\cal A}^\mu_a(k)=\left. g {\tilde J}^{\mu}_a(k)\right|_{k^2=0}$.  By decomposing the product $\epsilon_\mu(k) {\tilde\epsilon}_\nu(k)$ into its scalar and symmetric traceless graviton components\footnote{The anti-symmetric channel can be seen to vanish by explicit calculation.   On the gravity side, for spinless point sources this result can be understood from the symmetries of the two-form gauge field $B_{\mu\nu}$.}, the quantity ${\hat{\cal A}}^{\mu\nu}(k)$ was shown by explicit calculation~\cite{Goldberger:2016iau} to match the long distance radiation fields in a dilaton-graviton theory coupled to point particles, $S=S_g+S_{pp}$, with
\beq
\label{eq:gaction}
S_g= -2 m_{Pl}^{d-2}\int d^d x \sqrt{g} \left[R -(d-2) g^{\mu\nu}\partial_\mu\phi\partial_\nu\phi\right],
\eeq
and for a single particle $S_{pp} = -m\int d\tau e^\phi$.  In particular, the gravitational wave pattern measured by a detector at $r\rightarrow\infty$ (e.g. in $d=4$) 
\beq
h_{\pm}(t,{\vec n}) =  {4G_N\over r} \int {d\omega\over 2\pi} e^{-i\omega t} \epsilon^*{}_{\pm}^{ij}(k) {\tilde T}_{ij}(k),
\eeq
with $k^2=0$ is given by 
\beq
 -{1\over 2 m_{Pl}^{(d-2)/2}}\epsilon_{ij}(k) {\tilde T}^{ij}(k)=\epsilon_{ij}(k) {\hat{\cal A}}^{ij}(k),
\eeq
while the scalar radiation field is proportional to the Fourier transform of ${\cal A}_s(k)/r^{d-3}$ back to the time domain.

\section{Bi-adjoint radiation and its Yang-Mills double copy\label{sec:bi-adjoint}}

Our goal in this paper is to make contact between the classical Yang-Mills system of the previous section and its ``zeroth copy''~\cite{Luna:2015paa} obtained by demoting the gauge field $A^a_\mu(x)$ to a scalar $\phi^{a{\tilde a}}(x)$ transforming bi-linearly in the adjoint representation of two independent global symmetries $G$ and ${\tilde G}$.     This is motivated by results of~\cite{Luna:2015paa}, and by the work of~\cite{Cachazo:2013iea} which establishes a double copy relation between all tree-level amplitudes in $d$-dimensional pure Yang-Mills and those in the $G\times {\tilde G}$ invariant cubic scalar theory
\beq
\label{eq:bis}
{\cal L}_\phi= \frac{1}{2}(\del_\mu \p^{a {\tilde a}})^2 - \frac{y}{3} f^{abc} \ft^{{\tilde a} {\tilde b} {\tilde c}} \p_{a {\tilde a}} \p_{b {\tilde b}} \p_{c {\tilde c}}.
\eeq
Note that despite being allowed by symmetry in generic dimension $d$, a mass term as well as quartic and other self-interactions are not included, so at the quantum level this is a fine-tuned theory.   Because here we work purely at the classical level, we need not worry about the fact that loop corrections typically generate terms that would spoil the structure in Eq.~(\ref{eq:bis}).

In order to reproduce Eqs.~(\ref{eq:ym1}),~(\ref{eq:ym2}) as the double copy of an analogous classical solution in this bi-adjoint scalar field theory, we must introduce suitable point sources.   As in~\cite{Luna:2015paa}, we begin by identifying the static isolated Yang-Mills particle of momentum $p^\mu$ and charge $c^a$, whose Coulomb field is
\beq
A_\mu^a =   -g p_\mu c^a\int_\ell 2\pi\delta(\ell\cdot p) {e^{-i\ell\cdot x}\over\ell^2} 
\eeq
(in the gauge $\partial_\mu A^\mu_a=0$), with a bi-adjoint point charge corresponding to a static field configuration
\beq
\label{eq:sp}
\phi^{a\tilde a} =  - y c^a {\tilde c}^{\tilde a}\int_\ell 2\pi\delta(\ell\cdot p) {e^{-i\ell\cdot x}\over\ell^2}. 
\eeq
Thus, our system is described by an action $S=S_\phi+S_{pp}$, where for each point particle,
\beq
\label{eq:spp}
S_{pp} = -{1\over 2}\int d\lambda \left[\eta^{-1} {dx^\mu\over d\lambda} {dx_\mu\over d\lambda}  +\eta \left(m^2 - 2 y \phi^{a{\tilde a}} c_a {\tilde c}_{\tilde a}\right)\right].
\eeq
Here the einbein $\eta(\lambda)$ is a Lagrange multiplier, introduced to enforce invariance under reparametrizations $\lambda\rightarrow {\tilde\lambda}(\lambda)$ of the worldline time coordinate.  We have imposed by hand that the scalar couples universally to the point particles, with the same coupling strength $y$ as in the cubic self-interaction.  This particular choice of parameters seems to be yet another fine tuning of the theory, motivated only by the universal couplings of gluons to color charges in the putative Yang-Mills double copy.   While generic scalar worldline couplings such as the one in Eq.~(\ref{eq:spp}) normally get renormalized even at the classical level~\cite{Goldberger:2004jt,Porto:2007pw} by diagrams with insertions of the bulk interactions, the $G\times {\tilde G}$ color structure is such that in the present case, the relevant UV divergent diagrams vanish.   Thus, in the bi-adjoint theory symmetry protects the choice of worldline couplings from classical radiative corrections.

The equations of motion for the orbital coordinates $x^\mu(\lambda)$ are given by
\beq
{d p^\mu\over ds}  = -y c_a {\tilde c}_{\tilde a} \partial^\mu \phi^{a{\tilde a}},
\eeq
where the momentum is $p_\mu = -\delta S_{pp}/\delta {\dot x}^\mu=\eta^{-1} dx_\mu/d\lambda$ and $ds=\eta d\lambda$ is the re-parameterization invariant coordinate along the worldline.  Varying with respect to $\eta$ yields the on-shell constraint
\beq
p_\mu p^\mu - m^2 = -2y\phi^{a{\tilde a}} c_a {\tilde c}_{\tilde a}.
\eeq

In the point particle action Eq~(\ref{eq:spp}), we have not explicitly included terms whose variation yield the dynamics of the color charges $c_a, {\tilde c}_a$.  A model-independent approach to obtaining the equations of motion is to impose conservation of the currents $J^{\mu,a}$, $J^{\mu,\tilde a}$ that generate the $G\times {\tilde G}$ symmetry of the theory.   These currents can be decomposed as a sum
\beq
J^{\mu,a}=J^{\mu,a}_N + J^{\mu,a}_{pp}, \label{eq:baj}
\eeq
where $J^{\mu,a}_N= f^{abc}\phi^{c{\tilde b}} \partial^\mu\phi^{b{\tilde b}} $ is the Noether current implied by the invariance of ${\cal L}_\phi$, and to leading order in a derivative expansion
\beq
J^{\mu,a}_{pp} = \sum_\alpha \int ds p^\mu_\alpha c^a_\alpha(s) \delta^d(x-x_\alpha(s))+\cdots   \label{eq:bappj}
\eeq
is the current induced by the point charges.   The current $J^{\mu,\tilde a}$ that generates ${\tilde G}$ is defined analogously.   Then $\partial_\mu J^{\mu,a}=\partial_\mu J^{\mu,{\tilde a}}=0$ implies that the charges must evolve in time as
\begin{eqnarray}
{dc^a\over ds} =- y f^{abc} c^c {\tilde c}^{\tilde b}\phi^{b{\tilde b}}, & \displaystyle{{d{\tilde c}^a\over ds}}=- y {\tilde f}^{{\tilde a}{\tilde b}{\tilde c}} {\tilde c}^{\tilde c} {c}^{b}\phi^{b{\tilde b}},
\end{eqnarray}
after using the equations of motions for $\phi^{a{\tilde a}}(x)$ that result from varying $S_\phi+S_{pp}$. 

Alternatively, it is possible to derive the color equations of motion directly from the variation of a Lagrangian.   As a specific realization, one could introduce a worldline variable $\psi(\lambda)$ transforming bi-linearly in a representation $(r,{\tilde r})$ of $G\times {\tilde G}$.  Then, by defining the charges $c^a = \psi^\dagger \left(T^a_r\otimes {\mathbb I}_{\tilde r}\right) \psi,$ and  ${\tilde c}^{\tilde a} = \psi^\dagger \left(\mathbb{I}_r\otimes {\tilde T}^{\tilde a}_{\tilde r}\right) \psi,$ variation of the action
\beq
\label{eq:psis}
S_\psi =\int d\lambda \left[\psi^\dagger i\partial_\lambda\psi  + \eta y\phi^{a{\tilde a}} c_a {\tilde c}_{\tilde a} \right]
\eeq
yields the color equations of motion quoted above.  At the classical level, it is sufficient to work directly in terms of the equations of motion for the charges, so our results are independent of any choice of Lagrangian description such as Eq.~(\ref{eq:psis}).

To compare with Yang-Mills radiation, we construct the radiation field $\phi^{a\tilde a}$ measured by observers at $r\rightarrow\infty$.   The formal solution is expressible as 
\beq
\phi^{a{\tilde a}}(x) = - y \int_k \frac{e^{-i k \cdot x}}{k^2} \J^{a {\tilde a}}(k), 
\eeq
where the bi-adjoint source ${\cal J}^{a{\tilde a}}$ receives contributions from a collection of color charges, each coupled to $\phi^{a{\tilde a}}$ through the interaction in Eq.~(\ref{eq:spp}), as well as the scalar field configuration itself,
\beq
\J^{a{\tilde a}}(x) = -f^{abc} {\tilde f}^{{\tilde a}{\tilde b}{\tilde c}} \phi^{b{\tilde b}}\phi^{c{\tilde c}} +\sum_\alpha \int ds  c^a_\alpha(s)  {\tilde c}^{\tilde a}_\alpha(s) \delta^d(x-x_\alpha(s)).
\eeq
Like its Yang-Mills counterpart, this quantity has a simple relation to observables measured at asymptotic spatial distances $r=|{\vec x}|\rightarrow\infty$.   In particular, the long distance radiation field is (taking $d=4$ for illustration;   similar results hold in general dimension $d$)
\beq
\lim_{r\rightarrow\infty} \phi^{a{\tilde a}}(x) = {1\over 4\pi r} \int {d\omega\over 2\pi} e^{-i\omega t} {\cal A}^{a{\tilde a}}(k),
\eeq
where the amplitude ${\cal A}^{a{\tilde a}}(k) = y\left.\J^{a{\tilde a}}(k)\right|_{k^2=0}$ is evaluated at the on-shell momentum $k^\mu=\omega(1,{\vec x}/r)$, so that, for instance, the energy-momentum radiated out to infinity in the color channel $(a,{\tilde a})$ is given by
\beq
\Delta P^\mu_{a,\tilde a} = \int_k (2\pi)\theta(k^0)\delta(k^2) |{\cal A}^{a{\tilde a}}(k)|^2 k^\mu.
\eeq

In perturbation theory, as a formal expansion in the coupling $y$, the calculation of ${ \J}^{a{\tilde a}}(x)$ can be organized in terms of the Feynman diagrams shown in Fig.~\ref{fig:gluon1pt}, where now the wavy internal lines correspond to bi-adjoint scalar exchange.     As in the Yang-Mills case, we impose as initial conditions that the particles start out widely separated in the far past, with constant initial momenta $x^\mu_\alpha(s\rightarrow-\infty) = b^\mu_\alpha + p^\mu_\alpha s,$ and constant color factors $c^a_\alpha(s\rightarrow-\infty)= c^a_\alpha$, ${\tilde c}^{\tilde a}_\alpha(s\rightarrow-\infty)= {\tilde c}^{\tilde a}_\alpha$.   Thus, to leading order in perturbation theory the particles generate the static field given in Eq.~(\ref{eq:sp}) summed over all the particle sources,
\beq
\label{eq:lophi}
\phi^{a{\tilde a}}(x) = -y\sum_{\alpha} c_\alpha^a {\tilde c}_\alpha^{\tilde a}\int_\ell 2\pi\delta(\ell\cdot p_\alpha) {e^{-i\ell\cdot (x-b_\alpha)}\over\ell^2},
\eeq
and a radiation amplitude given by ${\cal A}^{a{\tilde a}}(k) =   y\sum_{\alpha} e^{ik\cdot b_\alpha} c^a_{\alpha} {\tilde c}^{\tilde a}_\alpha (2\pi) \delta(k\cdot p_\alpha)$.  This is non-vanishing only if $k^\mu$ points along the momentum of any massless particle source involved in the scattering process.

At the next order in the perturbative expansion, we need to determine how the leading order field in Eq.~(\ref{eq:lophi}) backreacts on the trajectories in orbital and color space.   We write
\beq
x^\mu_\alpha(s) = b^\mu_\alpha + p^\mu_\alpha s +z^\mu_\alpha(s),
\eeq
\begin{eqnarray}
c^a_\alpha(s) = c_\alpha^a + {\bar c}^a_\alpha(s), & \displaystyle{\tilde c}^{\tilde a}_\alpha(s) = {\tilde c}_\alpha^{\tilde a} + {\bar {\tilde c}}^{\tilde a}_\alpha(s),
\end{eqnarray}
with $z^\mu_{\alpha}(s), {\bar c}^a_\alpha(s),  {\bar {\tilde c}}^{\tilde a}_\alpha(s)$ vanishing as $s\rightarrow -\infty$.   Then, we feed the static field Eq.~(\ref{eq:lophi}) into the particle equations of motion, which yields the orbital deflection
\beq
\label{eq:z}
z_\alpha^\mu(\omega) = iy^2 \sum_{\beta\neq\alpha} (c_\alpha\cdot c_\beta)  ({\tilde c}_\alpha\cdot {\tilde c}_\beta) \int_\ell  {e^{-i\ell\cdot b_{\alpha\beta}}\over \ell^2 (\ell\cdot p_\alpha)^2} (2\pi)\delta(\ell\cdot p_\beta) (2\pi)\delta(\omega - \ell\cdot p_\alpha) \ell^\mu,
\eeq
as well as
\beq
\label{eq:c}
{\bar c}^a_\alpha(\omega) =- y^2 \sum_{\beta\neq\alpha}   \left[c_\alpha ,c_\beta\right]^a  ({\tilde c}_\alpha\cdot {\tilde c}_\beta)\int_\ell  {e^{-i\ell\cdot b_{\alpha\beta}}\over \ell^2 (\ell\cdot p_\alpha)} (2\pi)\delta(\ell\cdot p_\beta) (2\pi)\delta(\omega - \ell\cdot p_\alpha),
\eeq
and similarly for ${\bar {\tilde c}}^{\tilde a}_\alpha(s)$.    In these equations, we have defined Fourier transforms $z^\mu(\omega)=\int ds e^{is\omega} z^\mu(s)$ and ${\bar c}^a(\omega)=\int ds e^{is\omega} {\bar c}^a(s)$.

The ${\cal O}(y^2)$ radiation field receives contributions from Fig.~\ref{fig:gluon1pt}(a) evaluated using these time dependent deflections as well as from the diagram with the cubic vertex. In the latter case, it is sufficient to evaluate the diagram using the static worldlines, and we find a contribution to $\J^{a{\tilde a}}(k)$ which is given by
\beq
\label{eq:snlob}
\left. {\cal J}^{a{\tilde a}}(k)\right|_{\mbox{\footnotesize{Fig.}}~\ref{fig:gluon1pt}(b);{\cal O}(y^2)}=  y^2 \sum_{\alpha,\beta\atop \alpha\neq\beta} \int_{\ell_\alpha,\ell_\beta} \mu_{\alpha,\beta}(k)\left[c_\alpha, c_\beta\right]^a \left[{\tilde c}_\alpha,{\tilde c}_\beta\right]^{\tilde a} ,
\eeq
where $\mu_{\alpha\beta}(k)$ was defined in Eq.~(\ref{eq:mudef}).    The contribution to $\J^{a{\tilde a}}(k)$ from Fig.~\ref{fig:gluon1pt}(a),
\beq
\mbox{Fig.}~\ref{fig:gluon1pt}(a) = \sum_\alpha \int ds e^{ik\cdot x_\alpha(s)} c^a_\alpha(s) {\tilde c}_\alpha^{\tilde a}(s)
\eeq
becomes at ${\cal O}(y^2)$
\begin{eqnarray}
\label{eq:snloa}
\nonumber
\left. {\cal J}^{a{\tilde a}}(k)\right|_{\mbox{\footnotesize{Fig.}}~\ref{fig:gluon1pt}(a);{\cal O}(y^2)}
&=& -y^2\sum_{\alpha,\beta\atop \alpha\neq\beta}\int_{\ell_\alpha,\ell_\beta} \mu_{\alpha,\beta}(k) {\ell_\alpha^2\over k\cdot p_\alpha} \left[ {k\cdot\ell_\beta\over k\cdot p_\alpha} ((c_\alpha\cdot c_\beta) c_\alpha)^a (({\tilde c}_\alpha\cdot {\tilde c}_\beta){\tilde c}_{\alpha})^{\tilde a}\right.\\
& & \left. + \left[c_\alpha, c_\beta\right]^a\left(({\tilde c}_\alpha\cdot {\tilde c}_\beta) {\tilde c}_\alpha\right)^{\tilde a}
+ \left(({c}_\alpha\cdot { c}_\beta) c_\alpha\right)^a[{\tilde c}_\alpha, {\tilde c}_\beta]^{\tilde a} \right].
\end{eqnarray}
Here, the first term is due to the shift in the orbital trajectory Eq.~(\ref{eq:z}), while the second and third terms are the contributions of the color deflections ${\bar c}^a$ and ${\tilde {\bar c}}^a$ respectively.   The complete bi-adjoint current ${\cal J}^{a\tilde a}(k)$ at order $y^2$ is then the sum of of Eqs.~(\ref{eq:snlob}),~(\ref{eq:snloa}).

\subsection{Double copy}
\label{sec:dc}

We now apply the color-kinematics substitution rules in Eqs.~(\ref{eq:sub1}),~(\ref{eq:sub2}) to the bi-adjoint scalar amplitude determined above.   First, applying the replacement ${\tilde c}^a_\alpha\mapsto p_\alpha^\mu$ to the symmetry group ${\tilde G}$, yields the Yang-Mills current
\beq
{\cal J}^{a{\tilde a}}(k)\mapsto\sum_\alpha e^{ik\cdot b_\alpha} c^a_\alpha p^\mu_\alpha (2\pi)\delta(k\cdot p_\alpha) = {\tilde J}_a^\mu(k)   \label{eq:loymj}
\eeq    
at leading order in the couplings.   As discussed above, in sec.~\ref{sec:glue}, this expression implies vanishing gluon radiation at leading order in perturbation theory. 
 
In Yang-Mills, radiation first appears at second order in the gauge coupling.     By applying the replacement rule from Eq.~(\ref{eq:sub2})
\beq
[{\tilde c}_\alpha, {\tilde c}_\beta]^a\mapsto \Gamma^{\mu\nu\rho}(-k,\ell_\alpha,\ell_\beta) p_{\nu\alpha} p_{\rho\beta} = \left[(k\cdot p_\alpha) p_\beta^\mu - (k\cdot p_\beta) p_\alpha^\mu +{1\over 2} p_\alpha\cdot p_\beta (\ell_\alpha-\ell_\beta)^\mu\right]
\eeq
at the level of the integrand of the bi-adjoint solution at ${\cal O}(y^2)$, and using the constraints $\ell_\alpha\cdot p_\alpha=0$ from Eq.~(\ref{eq:mudef}), we arrive at
 \beq
\left. {\cal J}^{a{\tilde a}}(k)\right|_{\mbox{\footnotesize{Fig.}}~\ref{fig:gluon1pt}(b);{\cal O}(y^2)}\mapsto  -y^2 \sum_{\alpha,\beta\atop \alpha\neq\beta} \int_{\ell_\alpha,\ell_\beta} \mu_{\alpha,\beta}(k)\left[c_\alpha, c_\beta\right]^a \left(2(k\cdot p_\beta) p_\alpha^\mu - (p_\alpha\cdot p_\beta) \ell_\alpha^\mu\right),
\eeq
which has precisely the same structure as the Yang-Mills three-gluon contribution quoted in Eq.~(\ref{eq:ym2}). Similarly,
\begin{eqnarray}
\nonumber
\left. {\cal J}^{a{\tilde a}}(k)\right|_{\mbox{\footnotesize{Fig.}}~\ref{fig:gluon1pt}(a);{\cal O}(y^2)}&\mapsto& -y^2\sum_{\alpha,\beta\atop \alpha\neq\beta}\int_{\ell_\alpha,\ell_\beta} \mu_{\alpha,\beta}(k) {\ell_\alpha^2\over k\cdot p_\alpha} \left[((c_\alpha\cdot c_\beta)c_\alpha)^a \left((p_\alpha\cdot p_\beta) {k\cdot \ell_\beta\over k\cdot p_\alpha}p_\alpha^\mu + k\cdot p_\alpha p_\beta^\mu -  k\cdot p_\beta p_\alpha^\mu\right.\right.\\
& & \hspace{4.5cm}{}\left.  +\left.{1\over 2} p_\alpha\cdot p_\beta (\ell_\alpha-\ell_\beta)^\mu\right)+ [c_\alpha,c_\beta]^a (p_\alpha\cdot p_\beta) p_\alpha^\mu\right].
\end{eqnarray}
In this case, this expression differs from the integrand in Eq.~(\ref{eq:ym1}) by a term proportional to $k^\mu$, whose form is
\beq
\left[\sum_{\alpha,\beta\atop \alpha\neq\beta}\int_{\ell_\alpha,\ell_\beta} \mu_{\alpha,\beta}(k) ((c_\alpha\cdot c_\beta)c_\alpha)^a  {\ell_\alpha^2\over k\cdot p_\alpha}  (p_\alpha\cdot p_\beta) \right]k^\mu.
 \eeq
Thus, for off-shell $k^\mu$, the double copy of the bi-adjoint source ${\cal J}^{a{\tilde a}}(k)$ only reproduces the Yang-Mills current ${\tilde J}^\mu_a(k)$ up to gauge dependent terms.  However, the double copy of the bi-adjoint radiation field yields a gauge-invariant\footnote{Under gauge transformations that approach the identity at $|{\vec x}|\rightarrow\infty$.} on-shell gluon radiation field ${\hat{\cal A}}^a_\mu(k)$
\beq
{\cal A}^{a{\tilde a}}(k)\mapsto   \epsilon_\mu(k){\hat{\cal A}}_a^\mu(k),
\eeq
which is only defined up to ``pure gauge'' terms that vanish on-shell when dotted into the gluon polarization vector $\epsilon^\mu(k)$.    A similar gauge ambiguity arises in the double copy ${\cal A}^\mu_a(k)\mapsto {\hat{\cal A}}^{\mu\nu}(k)$ going from Yang-Mills to gravity~\cite{Goldberger:2016iau}.    This gauge freedom may be exploited to put the amplitude ${\hat{\cal A}}_a^\mu(k)$ into a form that automatically obeys the Ward identity $k_\mu {\hat {\cal A}}^\mu_a(k)=0$, 
\begin{eqnarray}
\nonumber
{\hat{\cal A}}_a^\mu(k) &=&  g^3\sum_{\alpha,\beta\atop \alpha\neq\beta} \int_{\ell_\alpha,\ell_\beta} \mu_{\alpha,\beta}(k)\left[ {\ell_\alpha^2\over k\cdot p_\alpha} ((c_\alpha\cdot c_\beta) c_\alpha)^a \left\{-p_\alpha\cdot p_\beta \left(\ell^\mu_\beta -{k\cdot\ell_\beta\over k\cdot p_\alpha} p^\mu_\alpha\right) +k\cdot p_\alpha p^\mu_\beta - k\cdot p_\beta p^\mu_\alpha\right\}\right.\\
& & \hspace{1cm} {}+\left. [c_\alpha,c_\beta]^a \left(2(k\cdot p_\beta) p^\mu_\alpha -  (p_\alpha\cdot p_\beta)\ell^\mu_\alpha+{\ell_\alpha^2\over k\cdot p_\alpha} (p_\alpha\cdot p_\beta) p_\alpha^\mu \right) \right].
\end{eqnarray}
After the identification $y\rightarrow -g$ in order to match normalization, this equation correctly reproduces the gluon radiation field detected by far away observers.

The color-kinematics substitution rules in Eqs.~(\ref{eq:sub1}),~(\ref{eq:sub2}) can also be applied to the conserved currents $J^{\mu,a}$, $J^{\mu,{\tilde a}}$ and energy-momentum tensor $T_\phi^{\mu\nu}$ of the scalar theory.   Evaluating these quantities on-shell (i.e. $\phi^{a{\tilde a}}$ and the particles obeying the equations of motion) then yields well defined double-copy currents
\begin{eqnarray}
J^{\mu,a}\mapsto {\hat J}^{\mu,a}, &  J^{\mu,{\tilde a}}\mapsto {\hat J}^{\mu\nu}, &   T_\phi^{\mu\nu}\mapsto {\hat T}^{\mu\nu}.
\end{eqnarray} 
We have explicitly constructed the currents ${\hat J}^{\mu,a},$ ${\hat J}^{\mu\nu},$ and ${\hat T}^{\mu\nu}$ and compared them to the analogous objects in the gauge theory setup of sec.~\ref{sec:glue}. The details of this calculation are presented in appendix \ref{sec:Noether}.   At leading order in perturbation theory, the currents only get contributions from the static point particle charges, and the mapping of currents is such that ${\hat J}^{\mu,a}$ coincides with the Yang-Mills color current, while both $ {\hat J}^{\mu\nu}$ and ${\hat T}^{\mu\nu}$ agree with the energy-momentum tensor of the color charges in gauge theory.   At second order, we find that ${\hat J}^{\mu,a}$ differs from the (gauge-dependent) Yang-Mills current ${\tilde J}^{\mu,a}$ defined in Eq.~(\ref{eq:gcurr})  by an improvement term of the form $\partial_\sigma A^{\sigma\mu,a}$, $A^{\sigma\mu,a}=-A^{\mu\sigma,a}$ and consequently leads to the same global charge $Q^a = \int d^{d-1} {\vec x} \,J^{0,a}(x)$ as one finds in the gauge theory.   The object ${\hat J}^{\mu\nu}$, which is the double copy of the ${\tilde G}$ global symmetry current, obeys the conservation law $\partial_\mu {\hat J}^{\mu\nu}=0$ and reproduces the total energy-momentum of the Yang-Mills system $P^\mu$, $\int d^{d-1} {\vec x}\, {\hat J}^{0\mu}(x)=P^\mu$, computed directly to order $g^2$ in Yang-Mills.  However, we find that ${\hat J}^{\mu\nu}(x)$ is neither gauge-invariant nor symmetric.    Finally ${\hat T}^{\mu\nu}(x)$ is symmetric and conserved, but not gauge invariant.    It also reproduces the conserved energy-momentum $P^\mu$ of the gauge theory\footnote{Of course, for our setup with point charges that are initially infinitely far apart, the global charges are simply $Q^a=\sum_\alpha c^a_\alpha$ and $P^\mu=\sum_\alpha p^\mu_\alpha$ to all orders in perturbation theory, so in this case we learn nothing from the fact that under the double-copy, the scalar theory charge $Q^a=\int d^{d-1} {\vec x} \,J^{0,a}\mapsto Q^a$ in Yang-Mills, or that ${\tilde Q}^{\tilde a}=\int d^{d-1} {\vec x} \,J^{0,{\tilde a}}(x)\mapsto P^\mu$.   Rather our explicit computation just serves as a consistency check of the color-kinematics substitution rules.   More generally, for instance in a configuration with bound particle orbits, the perturbative corrections to the global charges are non-trivial, and we expect the mapping of conserved quantities between bi-adjoint and Yang-Mills to have more physical content.}.

\section{Conclusions \label{sec:conclusion}}

In this paper, we have constructed perturbative radiating solutions of the bi-adjoint scalar theory coupled to dynamical point sources. Using the color-kinematics replacements summarized in Eqs.~(\ref{eq:sub1}), (\ref{eq:sub2}), we have found a double copy correspondence that relates the scalar field at asymptotic distances from the source to the analogous classical solution in gauge theory coupled to color charges.  The same replacement rules were found recently to relate gluon radiation to classical solutions in a theory of dilaton gravity. Thus the results of this paper imply a correspondence, at least at leading order in perturbation theory, between gravity and a scalar theory with much simpler Feynman rules.

We have focused here on the case of scattering solutions where the classical trajectories are unbound, but similar methods should apply to the case of non-relativistic bound orbits. Note that because the color-kinematics replacements are done at the level of the integrand, they relate perturbative solutions whose expansion parameters become small in different kinematic regimes.   For generic velocities and radiation frequency, the classical limit corresponds to large color charges  $c,{\tilde c}\gg 1$, while the double copy relations hold in the regime where $c,{\tilde c}$ are of order the orbital angular momentum $L\sim E b\gg \hbar$.    In the region of parameters $c/Eb,{\tilde c}/Eb\sim{\cal O}(1)$, the classical solutions in the bi-adjoint theory then correspond to an expansion in the dimensionless quantity $y^2  c \, {\tilde c} \, b^{5-d}/E\ll 1$ which controls the non-linearities as well as the corrections to the classical trajectories in orbital and color space.   The classical limit of the gauge theory is also $c\gg 1$, and in the regime of validity of the double copy there is a single expansion parameter of order $g^2 c \, b^{4-d}\ll 1$ that suppresses both the gluon self-interactions and the orbital and color deflections.    Finally,  in the gravitational double copy, the classical limit, with $E\gg m_{Pl}$, is perturbative as long as $E b^{3-d}/m_{Pl}^{d-2}\ll 1$, in which case both the gravitational non-linearities and the orbital deflections are under theoretical control.    

It remains to be seen if this method of obtaining gauge and gravity solutions by applying simple transformations holds beyond leading order in the interactions.   While the rules outlined in Eqs.~(\ref{eq:sub1}), (\ref{eq:sub2}) appear to be well defined at higher orders in perturbation theory, additional structure (such as the color and kinematic Jacobi identities that play a crucial role in BCJ duality for scattering amplitudes) may need to be imposed in order for the solutions to match.    Either a definite computation of the higher order corrections, or better, a general all-orders proof is needed\footnote{Recently~\cite{Cheung:2016say} constructed an action for $d$-dimensional gravity in which the interactions are expressed as a cubic structure with a manifest $SO(d-1,1)\times SO(d-1,1)$ Lorentz invariance.  One might speculate that this doubled kinematic invariance reflects an underlying $G\times{\tilde G}$ bi-adjoint structure, and it would be interesting to see whether constructing the gravitational solutions directly in the variables of~\cite{Cheung:2016say} could elucidate the connection with the bi-adjoint scalar to higher orders in the perturbative expansion.}.  Finally, even if the correspondence between bi-adjoint, gauge and gravity solutions holds to all orders, as alluded to in sec.~\ref{sec:intro}, a method of efficiently projecting out the dilaton (and at higher orders the two-form gauge field $B_{\mu\nu}$) to get radiation observables in pure gravity is still missing.    We hope to address these remaining issues in future work.

\section{Acknowledgments}

We thank Donal O'Connell for comments on the manuscript.   This research was partially supported by Department of Energy grant DE-FG02-92ER-40704.

\appendix

\section{Double copy of conserved currents \label{sec:Noether}} 

The double copy rules used to relate classical solutions can also be applied to the conserved currents of the scalar theory. These generate conserved currents whose physical content can be directly compared to $T^{\mu\nu}$ and the color current on the gauge theory side.  

First, we look at the $G$ global symmetry generating current of the bi-adjoint scalar theory,  $J^{\mu,a}(k)$ defined in Eq.~(\ref{eq:baj}).  At leading order in perturbation theory, the current is a sum of free particle terms
\beq
\left. J^{\mu,a}(k)\right|_{{\cal O}(y^0)} =  \sum_\alpha e^{i k 
\cdot b_\alpha} p^\mu_\alpha c^a_\alpha (2\pi) \delta(k \cdot p_\alpha).
\eeq
This expression is invariant under the double copy map and is identical to the Yang-Mills current at this order, Eq.~(\ref{eq:loymj}).  The leading order solutions for the field, the color variable and the trajectory are then inserted into Eq.~(\ref{eq:baj}) to obtain at ${\cal O}(y^2)$,
\beq
\left. J^{\mu,a}(k)\right|_{{\cal O}(y^2)}
= y^2\sum_{\alpha,\beta\atop \alpha\neq\beta}\int_{\ell_\alpha,\ell_\beta} \mu_{\alpha,\beta}(k) ({\tilde c}_\alpha\cdot {\tilde c}_\beta)  \left[ [c_\alpha,c_\beta]^a \left({\ell_\beta^\mu+ {\ell_\alpha^2\over k\cdot p_\alpha}}p_\alpha^\mu\right) - ((c_\alpha\cdot c_\beta) c_\alpha)^a  {\ell_\alpha^2\over k\cdot p_\alpha} \left({\ell_\beta^\mu- {k\cdot\ell_\beta\over k\cdot p_\alpha} p_\alpha^\mu }\right)  \right]. \label{eq:nloj}
\eeq
Applying the replacements ${\tilde c}^a_\alpha\mapsto p_\alpha^\mu$ and $y\mapsto-g$ maps this object onto a conserved current ${\hat J}^{\mu,a}(k)$ in the double copy gauge theory. We find that this differs from the (gauge dependent) Yang-Mills current $\tilde{J}^{\mu,a}(k)$ at this order (sum of Eqs.~(\ref{eq:ym1}),(\ref{eq:ym2})) by an improvement term of the form $k_\sigma M^{\sigma \mu,a}$, with 
\beq
M^{\mu \nu,a}(k)= - M^{\nu \mu,a}(k)= -g^2\sum_{\alpha,\beta\atop \alpha\neq\beta}\int_{\ell_\alpha,\ell_\beta} \mu_{\alpha,\beta}(k) \left[ [c_\alpha,c_\beta]^a - ( (c_\alpha\cdot c_\beta) c_\alpha)^a {\ell_\alpha^2\over k\cdot p_\alpha} \right] \left(p_\alpha^\mu p_\beta^\nu - p_\alpha^\nu p_\beta^\mu \right).
\eeq 
Consequently, ${\hat J}^{\mu,a}(k)$ is conserved and reproduces the same global charge as the Yang-Mills current $\tilde{J}^{\mu,a}(k)$ that sources the classical gauge field.

In the case of the energy-momentum tensor, there are two quantities in the bi-adjoint theory whose double copy can be compared to the gauge theory (symmetric, gauge invariant) $T^{\mu\nu}$:
\beq
\label{eq:siai}
T^{\mu \nu}=T^{\mu \nu}_{YM}+T^{\mu \nu}_{pp}=F_a^{\mu\lambda}F_a^{\nu}{}_\lambda-{1\over 4 }\eta^{\mu \nu}F_a^{\alpha\beta} F_{a\alpha\beta} + \sum_\alpha \int ds p^\mu_\alpha(s) p^\nu_\alpha(s) \delta^d(x-x_\alpha(s)).
\eeq 
One is the $G$ color current $J^{\mu,\tilde{a}}$ that maps onto an object ${\hat J}^{\mu\nu}$ under double copy.   The other is the scalar energy-momentum tensor itself $T_\phi^{\mu\nu}$ that can be interpreted as a quantity ${\hat T}^{\mu\nu}$ in gauge theory.    We find that these quantities obey conservation laws $\partial_\mu {\hat J}^{\mu\nu}=\partial_\mu {\hat T}^{\mu\nu}=0$ and generate the same global charge $P^\mu$ as the gauge theory $T^{\mu\nu}$, but differ from it locally.

To be explicit, on the gauge theory side, we have at leading order
\beq 
\left. T^{\mu\nu}(k)\right|_{{\cal O}(g^0)} =  \sum_\alpha e^{i k 
	\cdot b_\alpha} p^\mu_\alpha p^\nu_\alpha (2\pi) \delta(k \cdot p_\alpha)\label{eq:lot},
\eeq
which corresponds to free point particles.    At the next order in perturbation theory, the point-particle contribution is
\begin{eqnarray}
\left. T^{\mu \nu}_{pp}(k)\right|_{{\cal O}(g^2)}&=& -g^2 \sum_{\alpha,\beta\atop\alpha\neq\beta} \int_{\ell_\alpha,\ell_\beta}\mu_{\alpha,\beta}(k) (c_\alpha \cdot c_\beta) {\ell_\alpha^2\over k\cdot p_\alpha} \Big[ \left\{p_\alpha^\mu \left( (k\cdot p_\alpha)p_\beta^\nu-(p_\alpha\cdot p_\beta)\ell_\beta^\nu \right) +(\mu\leftrightarrow\nu) \right\}  \nonumber \\
& & \hspace{5.6cm}{}-  {p_\alpha^\mu p_\alpha^\nu \over k \cdot p_\alpha} \left( (k\cdot p_\alpha)(k\cdot p_\beta)-(p_\alpha\cdot p_\beta) (k\cdot \ell_\beta) \right) \Big], 
\end{eqnarray} 
whereas the bulk contribution is
\begin{eqnarray}
\left. T^{\mu \nu}_{YM}(k)\right|_{{\cal O}(g^2)}&=& g^2 \sum_{\alpha,\beta\atop\alpha\neq\beta} \int_{\ell_\alpha,\ell_\beta}\mu_{\alpha,\beta}(k) (c_\alpha \cdot c_\beta) \Big\{ \left[ \ell_\alpha^\mu p_\beta^\nu (k\cdot p_\alpha)+\ell_\beta^\nu p_\alpha^\mu (k\cdot p_\beta) -\ell_\alpha^\mu \ell_\beta^\nu (p_\alpha \cdot p_\beta)- p_\alpha^\mu p_\beta^\nu (\ell_\alpha \cdot \ell_\beta) \right]  \nonumber  \\ 
&& \hspace{4.5cm}{} +  {\eta^{\mu \nu}\over 2} \left[(\ell_\alpha \cdot \ell_\beta)(p_\alpha \cdot p_\beta)-(k\cdot p_\alpha)(k\cdot p_\beta) \right] \Big\}. 
\end{eqnarray} 
for on-shell fields.    

In the scalar theory, the current $J^{\mu,\tilde{a}}$ generating the $\tilde{G}$ global symmetry is also a sum of free particle currents to leading order
\beq
\left. J^{\mu,\tilde{a}}(k)\right|_{{\cal O}(y^0)} =  \sum_\alpha e^{i k 
	\cdot b_\alpha} p^\mu_\alpha \tilde{c}^{\tilde{a}}_\alpha (2\pi) \delta(k \cdot p_\alpha),
\eeq
and reproduces Eq.~(\ref{eq:lot}) upon replacing $\tilde{c}^{\tilde{a}}\rightarrow p^\mu$.   At the next order in perturbation theory,  $J^{\mu,\tilde{a}}$ is given by permutation of $G$ and ${\tilde G}$ labels in Eq.~(\ref{eq:nloj}), so under Eqs.~(\ref{eq:sub1}),~(\ref{eq:sub2}), we have $J^{\mu,\tilde{a}}\mapsto \hat{J}^{\mu \nu}$ which is given by
\begin{eqnarray}
\label{eq:jhat}
\left. \hat{J}^{\mu \nu}(k)\right|_{{\cal O}(g^2)}&=& -g^2 \sum_{\alpha,\beta\atop\alpha\neq\beta} \int_{\ell_\alpha,\ell_\beta}\mu_{\alpha,\beta}(k) (c_\alpha \cdot c_\beta)  \left[\left( \ell_\beta^\mu+ {\ell_\alpha^2\over k\cdot p_\alpha}p_\alpha^\mu \right)\left((k\cdot p_\alpha) p_\beta^\nu - (k\cdot p_\beta) p_\alpha^\nu +{1\over 2} p_\alpha\cdot p_\beta (\ell_\alpha-\ell_\beta)^\nu \right) \right. \nonumber \\
& & \hspace{4.8cm}{} \left. - {\ell_\alpha^2\over k\cdot p_\alpha} (p_\alpha\cdot p_\beta) \left(p_\alpha^\nu \ell_\beta^\mu - p_\alpha^\mu p_\alpha^\nu {k\cdot\ell_\beta\over k\cdot p_\alpha} \right) \right].
\end{eqnarray}
It can be checked that, if written in terms of the classical gauge field solutions $A^a_\mu(x)$, this expression is not invariant under gauge transformations $\delta A^a_\mu = D_\mu\alpha^a$.  Moreover, ${\hat J}_{\mu\nu}$ is not equivalent locally to the gauge invariant $T^{\mu\nu}$ in Eq.~(\ref{eq:lot}) by any simple improvement terms of the form $\partial_\sigma M^{\sigma;\mu\nu}$, $M^{\sigma;\mu\nu}=-M^{\mu;\sigma\nu}=-M^{\nu;\mu\sigma}$.   However, the global charge
 \begin{equation}
{\hat J}^\mu = \int  d^{d-1} {\vec x} \,\hat{J}^{0\nu}(x)= \int {dk^0\over 2\pi} e^{-ik^0 x^0} {\hat J}^{\mu\nu}(k^0,{\vec k}=0)
 \end{equation}
 computed from ${\hat J}^{\mu\nu}(k)$ does agree with the total-energy momentum $P^\mu$ of the gauge theory.

Similarly,  the scalar energy-momentum tensor $T^{\mu\nu}_\phi$ at leading order is identical to Eq.~(\ref{eq:lot}), whereas to order ${\cal O}(y^2)$, we get   
\begin{eqnarray}
\left. {T}^{\mu \nu}_\phi(k) \right|_{{\cal O}(y^2)}&=& y^2 \sum_{\alpha,\beta\atop\alpha\neq\beta} \int_{\ell_\alpha,\ell_\beta}\mu_{\alpha,\beta}(k) (c_\alpha \cdot c_\beta) (\tilde{c}_\alpha \cdot \tilde{c}_\beta) \left[{\ell_\alpha^2\over k\cdot p_\alpha}\left( \ell_\beta^\mu p_\alpha^\nu+ p_\alpha^\mu \ell_\beta^\nu -{k\cdot\ell_\beta\over k\cdot p_\alpha}p_\alpha^\mu p_\alpha^\nu \right) \right. \nonumber \\
& & \hspace{5.8cm}{} \left. - \ell_\alpha^\mu \ell_\beta^\nu + {\eta^{\mu \nu} \over 2} (\ell_\alpha \cdot \ell_\beta) \right]
\end{eqnarray}
The double copy ${\tilde c}^a_\alpha\mapsto p_\alpha^\mu$ then produces a well-defined current $\hat{T}^{\mu \nu}$ given by an expression similar to Eq.~(\ref{eq:jhat}).   It can be checked that, like ${\hat J}^{\mu\nu}$, this quantity while conserved and symmetric, is not invariant under gauge transformations $\delta A^a_\mu = D_\mu\alpha^a$ and cannot be related to $T^{\mu\nu}$ in Eq.~(\ref{eq:siai}) by adding improvement terms.    The global charge ${\hat T}^\mu=\int  d^{d-1} {\vec x} \,\hat{T}^{0,\nu}(x)$ is also in agreement with $P^\mu$ in the gauge theory.

The mismatch between the currents does not appear to be an impediment to the program of obtaining physical information about the gauge system from the scalar double copy.   The reason is that all the physical information about the gauge theory set-up (e.g, angular distributions of color, energy-momentum and angular momentum flux measured at asymptotic distances) can be expressed in terms of the quantity ${\cal A}(k)=\epsilon_\mu^a {\tilde J}^\mu_a(k)$ for $k^2=0$.   Because the double copy of the scalar solution reproduces ${\cal A}(k)$, it follows that the bi-adjoint theory encodes at least the observables of the gauge theory that can be accessed by measurements at infinity.

\end{document}